\documentclass[11pt,twoside]{article}
\usepackage{asp2010}

\resetcounters

\begin{document}

\title{The planetary nebula Abell 48 and its [WN4] central star}

\author{I.S. Boji\v{c}i\'c$^{1,2,3}$, D.J. Frew$^{1,2}$, Q.A. Parker$^{1,2,3}$, M. Stupar$^{1,2,3}$, S. Wachter$^{4}$ and K. DePew$^{1,2}$
\affil{$^1$Department of Physics and Astronomy, Macquarie University, NSW 2109, Australia}
\affil{$^2$Research Centre in Astronomy, Astrophysics and Astrophotonics, Macquarie University, NSW 2109, Australia}
\affil{$^3$Australian Astronomical Observatory, PO Box 915, North Ryde, NSW 1670, Australia}
\affil{$^4$Infrared Processing and Analysis Center, California Institute of Technology, Pasadena, CA 91125, USA}
}

\begin{abstract}
We have conducted a multi-wavelength study of the planetary nebula Abell~48 and give a revised classification of its nucleus as a hydrogen-deficient star of type [WN4].  The surrounding nebula has a morphology typical of PNe and importantly, is not enriched in nitrogen, and thus not the `peeled atmosphere' of a massive star.  Indeed, no WN4 star is known to be surrounded by such a compact nebula.  The ionized mass of the nebula is also a powerful discriminant between the low-mass PN and high-mass WR ejecta interpretations.  The ionized mass would be impossibly high if a distance corresponding to a Pop I star was adopted, but at a distance of 2~kpc, the mass is quite typical of moderately evolved PNe.  At this distance, the ionizing star then has a luminosity of $\sim$5000\,L$_{\odot}$, again rather typical for a PN central star.  We give a brief discussion of the implications of this discovery for the late-stage evolution of intermediate-mass stars.
\end{abstract}

\section{Introduction}
Planetary nebulae (PNe) are the short-lived shrouds of ionized gas ejected from AGB stars. However, recent discoveries are showing that the diversity of evolutionary pathways leading to PN formation is greater than previously believed \citep{wh06,FP10,FP12a,todt12} In the course of a spectroscopic survey of their central stars (CSPNe), we were struck by the unusual nature of the ionizing star of Abell 48.  
\citet{depew} gave a preliminary classification of its CSPN as [WN] or [WN/C], but since the important C\,{\small{IV}}~$\lambda$5806 doublet was unobserved, a more precise classification could not be given at that time.  Independently, 
\citet{wachter}  classified the central source as a massive WN6 star, in contrast to its previous long association as a PN. This uncertainty led us to more closely investigate this interesting object.  We have therefore conducted a detailed multi-wavelength study of Abell~48 and its central star, which confirms its PN nature and the [WN] status of its CSPN, as described below.

\section{Observations and Results}
We combine the spectrum of the CSPN taken with the Hale 200-inch telescope \citep{wachter} with spectra of both the nebula and central star taken with the WiFeS IFU on the ANU 2.3-m telescope at Siding Spring Observatory in 2009 and 2010. An analysis of these spectra shows that the CSPN is helium-rich (Figure 1). Based on the presence of strong N\,{\small{IV}} lines and moderately strong N\,{\small{V}} lines at $\lambda\lambda$4604,4620 relative to N\,{\small{III}} $\lambda\lambda$4634,4640, we classify it as [WN4] with an upper limit for hydrogen of 10\%. We also retrieved archival multi-wavelength data via the CDS for the star and nebula.  From the CSPN colours and nebular Balmer decrement, we determine $E(B-V)$ = 2.0. Adopting this reddening and the integrated nebular flux from \citet{FIP12},  log\,$F$(H$\alpha$) = $-11.5$, we estimate a distance of 2.0\,kpc, giving the CSPN a luminosity of $\sim$5000\,L$_{\odot}$.

\begin{figure}
\begin{center}
\includegraphics[width=9cm]{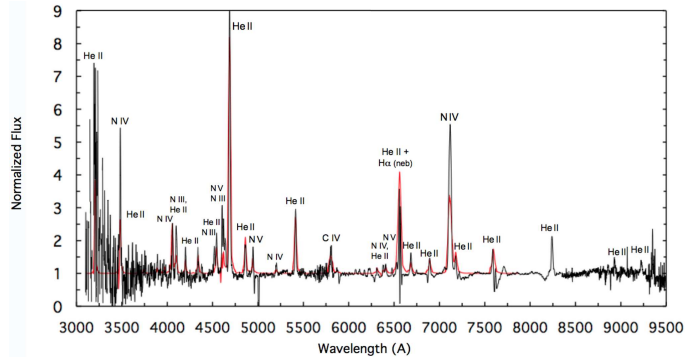}
\caption{Normalized spectrum (black line) of the central star.  Overplotted is model 10-13 (red line) from the Potsdam Wolf-Rayet grid (PoWR; \citealt{hamann2004}), which fits the relative strengths of the He lines well. The model parameters are $T_{\rm eff}$ = 71kK and log$R_t$ = 0.8, with the abundances set to He:C:N:O = 0.98: 1E-4: 0.015: 0.0.  The relatively poor fit to the nitrogen lines suggests the adopted N abundance is at least a factor of two too low, or 3 to 4\%.}
\label{fig:CSPN_spec}
\end{center}
\end{figure}

\section{The CSPN of Abell~48: unambiguously not a massive star}
\begin{figure}
\begin{center}
\includegraphics[width=4.6cm]{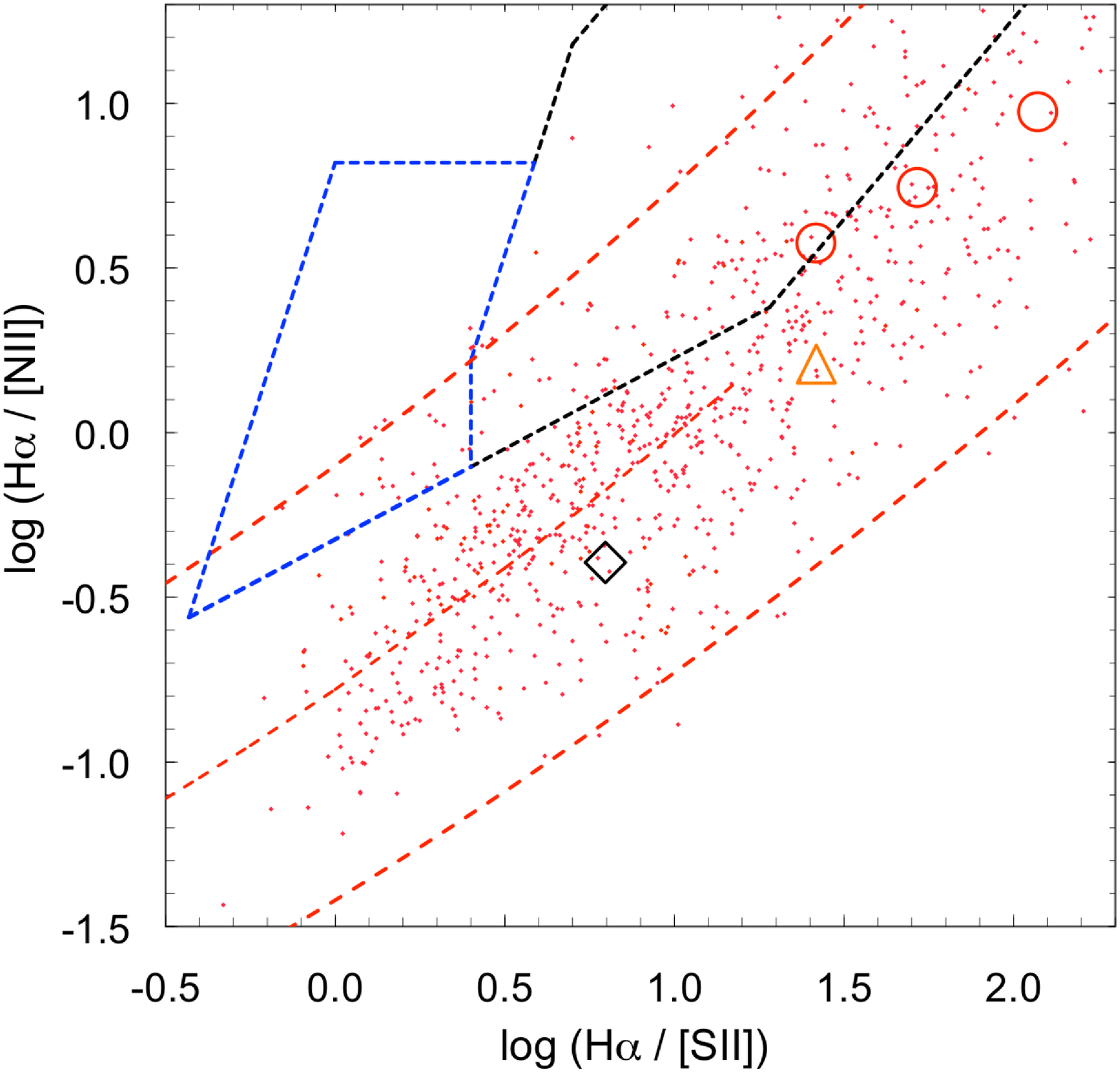}
\includegraphics[width=4.6cm]{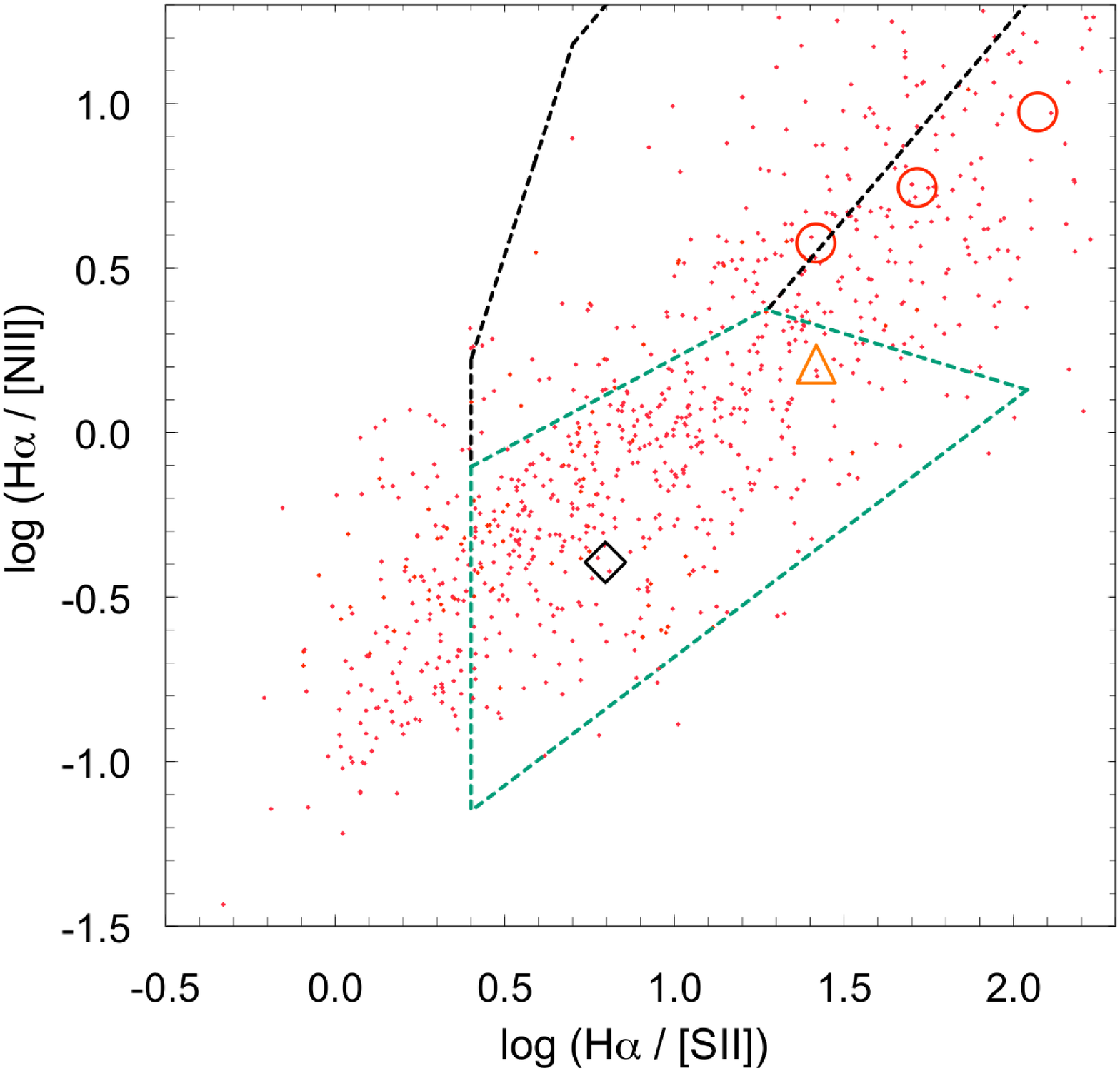}
\caption{Diagnostic plots updated from \citep{FP10}.  Individual PNe are shown as small dots and fields showing HII regions, LBV/WR ejecta, SNRs and PNe are marked; note the considerable overlap between them.  The red circles show Abell~48, IC~4663, and PB~8  (refer to the text), which all plot squarely in the PN domain, while the triangle is LMC N66.  On the other hand, the ring nebula around the WN7b star PMR~5 (black diamond) is strongly nitrogen enriched from CNO cycling, and plots in the area which includes both LBV/WNL ejecta and Type\,I PNe.}
\label{fig:diagnostic}
\end{center}
\end{figure}

\citet{misza} dismissed Abell~48 as a massive star, but gave no justification for this conclusion.  Based on an objective assessment of all currently available multi-wavelength data, we discount a Population I interpretation, and conclude that this is a bona fide PN.  Critically, the surrounding nebula is not enriched in nitrogen, indicating it is not the `peeled atmosphere' of a massive star.  Indeed, no WN4 star is known to be surrounded by such a compact nebula.Ê Furthermore, given the known reddening and  adopting an appropriate absolute magnitude from \citet{crowther}, it would be at a distance of $>$11\,kpc if it was truly a massive star.  This location is on the far side of the Galactic bar, and cannot be reconciled with the observed reddening, a point alluded to by \citet{wachter}.  The nebular ionized mass (determined from the distance, nebular diameter, H$\alpha$ flux, and reddening) would also be impossibly high, around 40\,$M_{\odot}$.  In addition, the morphology of the nebula and its location in the line-H$\alpha$/[NII] vs H$\alpha$/[SII] diagnostic plot (Fig.~\ref{fig:diagnostic}), indicate that Abell~48 is a bona fide PN.  While any identification as a massive star is highly problematic, we emphasize that {\it there are no observables in conflict with the PN interpretation}.

\section{Related Objects}
Unlike the peculiar nebula LMC N66, which may be a nebula around an accreting binary star \citep{hamann2003}, we find no evidence of spectroscopic or photometric variability in the CSPN of Abell~48.  The atmospheric composition is also different to the [WN/WC] nucleus of PB 8 \citep{todt10a,mb11}, illustrating the diversity of compositions seen in post-AGB stars.   After we commenced this study, \citet{misza} discovered a hot, helium-rich [WN3] CSPN in IC~4663, with $T_{\rm eff}$ = 140\,kK and an estimated luminosity of 4000\,L$_{\odot}$, quite typical of a CSPN nearing the knee in its evolutionary track.   Incidentally, we also reclassified PMR~5 \citep{mpc2003,todt10b} on the basis of a deeper WiFeS spectrum, and revise its classification to WN7b.  We also show that the surrounding nebula is composed of strongly CNO-processed material ejected from a massive star at a distance of about 3.5\,kpc.  A fuller account will be published separately.

\begin{figure}
\begin{center}
\includegraphics[width=13cm]{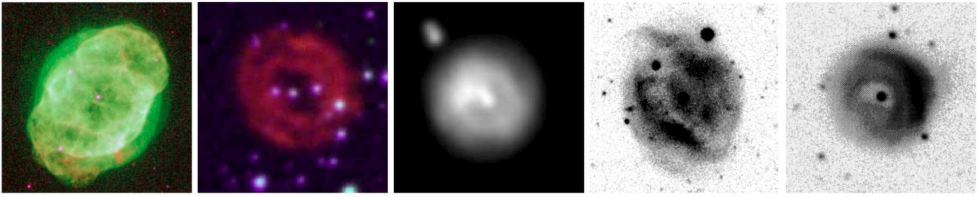}
\caption{Image montage of IC~4663, Abell~48, PB~8, K~1-27, and LoTr~4 respectively (images from  \citealt{hajian}, \citealt{scm92} and \citealt{rauch96,rauch98}). The  morphology of Abell 48 is quite similar to K~1-27 and LoTr4.}
\label{fig:montage}
\end{center}
\end{figure}

\section{Discussion and Future Work}
The recent discovery of two helium-dominated Galactic [WN] CSPN by \citet[Abell~48]{depew} and \citet[IC~4663]{misza}  
 has confirmed that the diversity of post-AGB pathways is greater than previously thought.  The [WN] stars are the progenitors of the O(He) stars \citep{rauch94,rauch98,rauch06,reindl,misza}, and will ultimately become DO white dwarfs as they fade and cool.  However their immediate progenitors remain uncertain, but are possibly the R CrB stars (e.g. \citealt{rauch08}).  There is an early suggestion that the scale height of Galactic  [WN/C] stars and their offspring, the O(He) stars, is larger than for the PN population as a whole \citep{FP12a}, which suggests that these objects may derive from lower-mass progenitor stars.  To test this hypothesis we are obtaining new observations of additional candidate [WN] or [WN/C] stars (e.g. \citealt{pm2003,todt12}). Clearly larger samples are required before definite conclusions can be made on their origin, so much more work needs to be done.  


\bibliography{Bojicic_A48rev}

\end{document}